\documentstyle[preprint,aps]{revtex}

\newcommand{\rv}{{\bf r}}

\newcommand{\ev}{{\bf e}}
\newcommand{\Ev}{{\bf E}}

\newcommand{\dv}{{\bf d}}

\newcommand{\Dcv}{\hbox{\boldmath$\cal D$}}

\newcommand{\kv}{{\bf k}}

\newcommand{\beq}{\begin{equation}}
\newcommand{\eeq}{\end{equation}}
\newcommand{\bea}{\begin{eqnarray}}
\newcommand{\eea}{\end{eqnarray}}

\begin{document}
\draft
\preprint{}
\title{Spontaneous photon emission stimulated \\
by two Bose condensates}
\author{ C.M. Savage \cite{craigaddress}, Janne Ruostekoski and Dan 
F. Walls}
\address{
Department of Physics, University of Auckland, Private Bag 92019, 
\\Auckland,
New Zealand}
\date{\today}
\maketitle
\begin{abstract}
We show that the phase difference of two overlapping ground state 
Bose-Einstein condensates can effect the optical spontaneous emission 
rate of excited atoms.  Depending on the phase difference the atom 
stimulated spontaneous emission rate can vary between zero and the 
rate corresponding to all the ground state atoms in a single 
condensate.  Besides giving control over spontaneous emission this 
provides an optical method for detecting the condensate phase 
difference.  It differs from previous methods in that no light fields 
are applied.  Instead the light is spontaneously emitted when excited 
atoms make a transition into either condensate.
\end{abstract}

Spontaneous symmetry breaking is a fundamental physical phenomenon.  
In particular, breaking of the global phase symmetry explains much of 
the interesting physics of Bose-Einstein condensates \cite{FOR75}.  We 
show that this spontaneously broken symmetry can effect the rate of 
spontaneous emission from excited atoms.  Hence we are able to link 
two fundamental physical phenomena: spontaneous symmetry breaking and 
spontaneous emission.  

Other physics which effects the spontaneous emission rate are; atom 
stimulation by a single condensate \cite{Hopesecs}, non-free space 
electromagnetic boundary conditions \cite{API}, due to a mirror for 
example, and superradiant type dipole correlations \cite{API}.  The 
latter occurs when the atoms are much closer than the wavelength of 
the emitted light.  Our effect is different to these and happens when 
the excited atoms can decay into either of two final states which are 
Bose condensed.  Its physical origin is the indistinguishability of 
the final states, which produces phase dependent interference terms in 
the transition probability.

Two groups have reported producing two dilute gas Bose-Einstein 
condensates in the same trap.  Spatially overlapping condensates of 
the $|F=1,m=-1 \rangle$ and $|F=2,m=2 \rangle$ hyperfine spin states 
of rubidium-87 have been produced by sympathetic cooling \cite{Myatt}.  
Separated condensates of the same state of sodium-23 have been 
produced using a far blue detuned laser sheet to divide a magnetic 
trap \cite{Andrews}.  Our system requires two overlapping condensates 
into which an excited state can make a spontaneous transition.

Bose stimulation of transitions by {\em atoms} has been 
studied for both electronic \cite{Hopesecs} and nuclear 
\cite{Namiot,Hoperadsecs} transitions. It was shown to be 
experimentally feasible for electronic transitions \cite{Hopesecs}. 
We apply the same idea to a system with two ground 
state condensates instead of one, and find that the atom stimulated 
emission rate depends on the condensate phase difference.

This phase difference has been observed directly by destructive 
optical imaging of the interference fringes in the density of 
overlapping condensates \cite{Andrews}.  Previously proposed methods 
for optically detecting this phase difference require driving Raman 
transitions between the ground states.  The excited state was created 
only through coherent excitation of the ground states and hence was 
not independent of the ground states.  Javanainen has shown that 
prompt amplification of one Raman beam by light-stimulated scattering 
is a signature of a condensate phase difference \cite{JAV96b}.  This 
is because the condensate phase difference behaves like a Raman 
optical coherence.  Imamo\=glu and Kennedy have studied light 
scattering from two spatially separated condensates.  They showed that 
the optical coherence allows spontaneous Raman scattering to generate 
coherent light, as well as allowing the complete elimination of Raman 
scattering \cite{IMA96}.  These two schemes were treated 
qualitatively, in contrast to the fully quantitative treatment we 
shall give.  Ruostekoski and Walls found that the relative peak 
heights of the characteristic two-peaked incoherent light scattering 
spectrum depend on the relative condensate phase \cite{Ruostekoski96}.  
Our method is related to these three insofar as it results from the 
macroscopic quantum coherence between the two condensates.  However no 
external light is applied and no Raman transitions occur.  The light 
is emitted by the decay of independently produced excited atoms.  A 
schematic experimental setup is shown in Fig.  1.

We consider transitions from an excited state labeled by $e$ to ground 
states labeled by $g_1$ and $g_2$.  The excited state is separately 
prepared and then launched towards the ground state condensates.
The ground states may differ in their 
internal quantum numbers and/or in their external wavefunctions.  Both 
situations have been achieved experimentally \cite{Myatt,Andrews}.
We neglect radiation reaction and multiple scattering 
effects \cite{Morice,Ruostekoski97}.  This requires that the system be 
optically thin at the frequency of the spontaneously emitted light 
\cite{JAV95b}.  We show that the recoil shift of the emitted photon 
makes this possible, provided the natural linewidth is sufficiently 
small.  Hence we have in mind the decay of a metastable excited state.

We consider spontaneous emission from excited atoms.
``Spontaneous'' specifies that the emission is not stimulated by 
light, although it may be stimulated by atoms.  The spontaneous 
emission is from the excited state atom field $\hat{\psi}_{e}$ to  
ground state atom fields $\hat{\psi}_{g_{i}}$.  The positive frequency part 
of the electric field operator at position $\rv$ is given in terms of 
the atom fields by Javanainen and Ruostekoski \cite{JAV95b} as 
\beq  
\hat{\Ev}^{+} (\rv) = \hat{\Ev}^{+}_{F} (\rv) +
\sum_{g} \int d^3 \rv' \;
{\bf K} ( \dv_{g} ; \rv, \rv') 
\hat{\psi}_{g}^{\dagger} (\rv') \hat{\psi}_{e} (\rv') ,
\label{electric field}
\eeq
where $\hat{\Ev}^{+}_{F} (\rv)$ is the free field, which we assume to 
be the vacuum so that $\langle \hat{\Ev}^{-}_{F} (\rv) \cdot
\hat{\Ev}^{+}_{F} (\rv) \rangle = 0$.  Rotation at the transition 
frequency $\Omega$ has been removed from the excited atom and electric 
fields, so that they are slowly varying \cite{JAV95b}.  We ignore 
excited state depletion and ground state growth.  The integration 
kernel ${\bf K} ( \dv ; \rv, \rv') $ generates the field at $\rv$ due 
to a dipole moment $\dv$ at $\rv'$ \cite{JAC75}.  We shall only be 
concerned with electromagnetic waves in the far field so that the 
dominant part of the kernel is that proportional to $| \rv -\rv' 
|^{-1}$, 
\beq
{\bf K} ( \dv_{g} ; \rv, \rv') \approx
\frac{k^2}{4 \pi \varepsilon_0} 
\frac{\exp(i k |\rv|)}{|\rv|}
\exp(-i k \rv' \cdot {\bf n} )
\Dcv_{g} ,
\label{kernel}
\eeq
where $\dv_g$ is the dipole moment of the transition $g 
\leftrightarrow e$, and $k=|\kv|$ where $\kv$ is the wavevector of the 
emitted light.  The far field assumption, in the form $|\rv| \gg 
|\rv'|$, has been used to replace $| \rv -\rv' |$ in the denominator 
by $|\rv|$ and to expand the exponential $\exp(ik|\rv -\rv'|) \approx 
\exp(ik|\rv| -ik \rv' \cdot {\bf n})$, where ${\bf n} = \rv / |\rv|$.  
We have also approximated the propagation direction $(\rv -\rv') / 
|\rv -\rv'|$ by ${\bf n}$, and hence introduced the components of the 
dipole moments perpendicular to the propagation direction ${\bf n}$, 
\beq
\Dcv_g = ( {\bf n} \times \dv_g ) \times {\bf n} 
= \dv_g -{\bf n} ({\bf n} \cdot \dv_g) .
\label{defE}
\eeq

With these approximations the expectation value of the field 
intensity, with respect to the atom field states and a vacuum free 
field, is 
\bea
I (\rv) &=&  2c \varepsilon_0 
\langle \hat{\Ev}^{-} (\rv) \cdot \hat{\Ev}^{+} (\rv) \rangle
\nonumber \\
&=& \frac{\kappa}{|\rv|^{2}} \sum_{g,g'} \int d^{3} \rv' \int d^{3} \rv''
\exp[-ik (\rv'' -\rv') \cdot {\bf n} ] \times
\nonumber \\
&& \Dcv_{g}^* \cdot \Dcv_{g'} 
\langle \hat{\psi}_{e}^{\dagger} (\rv') \hat{\psi}_{g} (\rv')
\hat{\psi}_{g'}^{\dagger} (\rv'') \hat{\psi}_{e} (\rv'')
\rangle ,
\label{intensity}
\eea
where $\kappa = c k^4 / (8 \pi^{2} \varepsilon_0 ) $.  As 
usual we consider the atom field operators to be expanded in 
complete sets of orthonormal mode functions $\{ \phi_{i} (\rv) \}$ 
which are adapted to the system under consideration 
$\hat{\psi} (\rv) = \sum_{i} \hat{a}_{i} \phi_{i} (\rv)$, where the 
$\hat{a}_{i}$ are the usual mode annihilation operators.  
The ground state modes into which condensation occurs are typically 
solutions to the Gross-Pitaevskii equation \cite{Lifshitz}.

We assume a sufficiently low temperature that stimulation by the 
non-condensed atoms can be ignored. In a JILA experiment up to 80\%
of the trapped atoms have been measured to be condensed, at a 
temperature of about 140 nK \cite{Ensher}.  Since the ratio of 
non-condensed to condensed atoms is then $1/4$ this implies that 
spontaneous emission stimulated by non-condensed atoms can be reduced, 
at least, to $1/4$ of that stimulated by the condensates.  This would 
be a constant background emission.  It would add to the unstimulated 
free space spontaneous emission rate $\gamma$, which is {\it always} 
present.  Furthermore, in their two condensate experiment the JILA 
group reports cooling proceeding until ``we can no longer see any 
noncondensed atoms'' \cite{Myatt}.

We assume the Bose condensed modes may be represented by coherent 
states $| \sqrt{N} \exp(i \theta) \rangle$.  The coherent state phase 
$\theta$ embodies the broken global phase symmetry.  Since 
Hamiltonians always contain pairs of atom creation and annihilation 
field operators the phase is unmeasureable.  However phase differences 
between condensates are measureable, although experimentally they are 
not yet controllable and hence they vary from shot to shot.  The 
coherent states have a mean atom number of $N$ and an atom number 
uncertainty of $\sqrt{N}$.  We further assume that all condensates are 
initially uncorrelated.  The assumption that the excited state is 
condensed is not essential, however it does simplify our treatment.  
The atom field operator expectation value determining the intensity 
Eq.(\ref{intensity}) is then 
\beq
\langle \hat{\psi}_{e}^{\dagger} (\rv') \hat{\psi}_{g} (\rv')
\hat{\psi}_{g'}^{\dagger} (\rv'') \hat{\psi}_{e} (\rv'')
\rangle =
N_e \sqrt{N_{g} N_{g'}} \exp(i \delta \theta_{gg'})
\phi_{e}^{*} (\rv') 
\phi_{g} (\rv') \phi_{g'}^{*} (\rv'') \phi_{e} (\rv'') ,
\label{intensity4op}
\eeq
where $\delta \theta_{gg'} = \theta_g -\theta_{g'}$ is the ground 
state condensate phase difference. The atom stimulated intensity 
can then be written as
\beq
I (\rv) = \frac{\kappa}{|\rv|^2} N_e \left\{
\sum_{g} N_g | C_{g} \Dcv_g |^2 +
2 \sqrt{N_{g_{1}} N_{g_{2}}} \mbox{Re} \left\{
\exp(i \delta \theta_{g_1 g_2})
\Dcv_{g_1}^* \cdot \Dcv_{g_2} C_{g_{1}}^{*} C_{g_{2}} \right\}
\right\} ,
\label{intensity nume}
\eeq
where $\mbox{Re}\{\}$ means the real part, and we have defined 
Franck-Condon factors, or the 
overlaps of the excited and ground modes taking into account the 
photon recoil,
\beq
C_{g} = \int d^{3} \rv'  
\exp(-ik \rv' \cdot {\bf n} ) 
\phi_{g}^{*} (\rv') \phi_{e} (\rv') .
\label{wavefunction overlap}
\eeq
The last term in the expression for the intensity Eq.(\ref{intensity 
nume}) depends on the condensate phase difference $\delta 
\theta_{g_{1} g_{2}}$ provided the Franck-Condon factors are non-zero 
and $\Dcv_{g_1}^* \cdot \Dcv_{g_2} \neq 0$. 
That is provided there is some field polarization that {\em both} 
dipole moments can generate at $\rv$.  Since the 
electric field is a vector field this is necessary for interference to 
be possible.  This dependence of the spontaneously emitted intensity on 
the condensate phase difference could in principle be used to measure 
it.

We now give a physical explanation based on the quantum mechanical 
transition amplitudes.  First we recall that if the final states of 
two transitions are in principle distinguishable then the individual 
transition probabilities must be added to get the final state 
probability.  However if the final states are indistinguishable then 
the transition {\em amplitudes} must be added, before taking the 
squared modulus to get the final state probability.  The squaring 
produces interference terms in the final state probability.  The 
intensity radiated in the propagation direction ${\bf n}$ is 
proportional to the transition probability for producing a photon 
propagating in that direction.  The first terms in Eq.(\ref{intensity 
nume}) correspond to the individual probabilities for the transitions 
into each ground state.  The last term corresponds to the product of 
amplitudes for each transition and hence represents quantum mechanical 
interference between them.  Interference occurs because the condition 
$\Dcv_{g_1}^* \cdot \Dcv_{g_2} \neq 0$ ensures that the particular 
transition which produces photons with certain polarizations cannot be 
distinguished.  A general treatment of the electric field polarization 
dependence on the atomic level scheme is given by Javanainen and 
Ruostekoski \cite{JAV95b}.

We next estimate the mode overlaps $C_g$.  We use gaussians for the 
ground mode functions
\beq
\phi_g = (2 \pi l^2 )^{-3/4} 
\exp[ - ( \rv \mp D \ev_{z} )^2 / (4 l^2) ] ,
\label{wf}
\eeq
where $l^2$ is the one dimensional variance of the probability density 
$\phi_{g}^2$, $\ev_{z}$ is the unit vector in the $z$ direction, and 
we have allowed for possible displacement by the distance $\pm D$ in 
this direction.  For the excited atoms we use the plane wave mode 
$\phi_e = V^{-1/2} \exp(i k_e x)$.  This represents atoms propagating 
in the positive $x$ direction with momentum $\hbar k_e$.  $V$ is the 
mode volume.  The plane wave usefully models modes which are much 
larger than the ground modes, so for purely calculational reasons we 
require $(2l)^{3} / V \ll 1$.  We assume that the excited state atoms 
are completely independent of the ground state atoms. They would be 
prepared in a separate part of the experiment.
Evaluating the integral 
Eq.(\ref{wavefunction overlap}) we find 
\bea
C_g &=&  ( 8 \pi )^{3/4}  
\left( \frac{l^3}{ V } \right)^{1/2} 
\exp[ -l^2 ( k^2 +k_{e}^{2} -2k k_e \cos \vartheta ) ] \times 
\nonumber \\ 
&& \exp[ \pm i k D \sin \vartheta \cos \varphi ] ,
\label{wf overlap example}
\eea
where $\vartheta$ is the angle between the light emission direction 
${\bf n}$ and the excited state propagation direction $\ev_{x}$, and 
$\varphi$ is the polar angle about $\ev_{x}$.  The first exponential 
results from momentum conservation while the second is a phase 
modulation due to displacement of the condensate from the coordinate 
origin. 

Since the mode functions $\{ \phi_{i} \}$ are a complete orthonormal 
set the fraction of emissions into the solid angle $d \Omega$ with 
final atomic mode $\phi_g$ is $|C_g|^2 \; d \Omega$.  Choosing the 
photon and excited state momenta equal $k = k_e$, which gives the 
optimal overlap, and using the realistic values $l^3 / V = 0.01$, and 
$l k = 100$ gives 
\bea
|C_g|^2 &\approx& 1.3 
\exp[ -4 \times 10^4 ( 1 - \cos \vartheta) ]
\nonumber \\
&\approx& 1.3 \exp[ - 2 \times 10^4 \vartheta^2 ], \; \; \;
( \vartheta \ll 1 )  .
\label{C numerical}
\eea
This mode overlap is largest for emission into a cone about the 
excited state propagation direction \cite{Hopesecs}.
For the atom stimulated emission rate to exceed the non-stimulated 
rate by a factor of one hundred, at an emission angle of
$\vartheta = 0.02$ radians, requires about $100/|C_g|^2 \approx 2.3 
\times 10^{5}$ atoms in the stimulating condensate. At $\vartheta 
=0.03$ radians the stimulated emission is negligible.

A condition for the validity of our analysis is that the spontaneously 
emitted light is sufficiently far off resonance to avoid cooperative 
effects and multiple scattering \cite{JAV95b}.  Assuming $k = k_{e}$, 
the emitted photon is detuned from resonance by the recoil frequency 
$\omega_{R} = \hbar k^{2} /(2m)$, where $m$ is the atomic mass.  This 
detuning must be at least of the order of the condensate collective 
linewidth, which has been estimated to be $3N_{g} \gamma /(2 
(lk)^{2})$, where $\gamma$ is the transition's free space natural 
linewidth \cite{JAV94,You}.  The condition is then 
\beq
\omega_{R} > \frac{3}{2} \frac{N_{g} \gamma}{(lk)^{2}} .
\eeq
With $N_{g} =10^{6}$ and $lk=100$ this becomes $\omega_{R} > 150 
\gamma$.  A typical recoil frequency of $\omega_{R} = 10^{5}$/s 
requires the natural linewith $\gamma < 600$/s.  We note that the 
total transition rate is proportional to $N_{e}$ which is not 
constrained.

We next consider some particular examples.  Let the excited state have 
magnetic quantum number $m$.  It can decay by emission of   
$(\sigma^-)$ or $(\sigma^+)$ circularly polarised photons, or by 
emission of a linearly polarised photon.  The respective final states 
having magnetic quantum numbers $m+1$, $m-1$ and $m$.  We assume that 
transitions into the $m$ ground state, which is not condensed,
can be ignored.  This is 
justified when transitions into the condensed $m \pm 1$ modes are 
dominant because of Bose stimulation.

The total radiated power is found by integrating Eq.(\ref{intensity 
nume}), for the intensity radiated in direction ${\bf n}$, over all 
directions.  Let $z$ be the quantization axis and let the excited 
atoms propagate in the positive $x$ direction, see Fig.  1.  We use 
the circular dipole moments $\dv_{\pm} = \mp d_{\pm} (\ev_x \pm i 
\ev_y)/\sqrt{2}$, where $d_{\pm}=|\dv_{\pm}|$ are the magnitudes of 
the dipole moments, and the $\ev_{x/y}$ are unit vectors in the $x/y$ 
directions.  In the $z$ direction these dipoles radiate photons with 
opposite circular polarization. In the $x/y$ plane, however, they 
radiate indistinguishable linearly polarized photons, which makes 
interference possible, see Fig. 1.
Using the previously calculated mode overlaps Eq.(\ref{wf 
overlap example}) the total atom stimulated radiated power is found to 
be 
\bea
P &=& \int I(\rv) |\rv|^2 d^2 {\bf n} 
\nonumber \\
&=& A N_{e} \left\{ B ( N_+ d_{+}^{2} +N_- d_{-}^{2} ) +
2 C \sqrt{N_+ N_-} d_+ d_- \cos \delta \theta_{+-} \right\} .
\label{power}
\eea
The integral is over the spherical surface of radius $|\rv|$ centered on 
the origin.  We have defined the coefficients 
\bea
A &=& \kappa ( 8 \pi )^{3/2}  
\frac{l^3}{ V } 2 \pi
\exp[ -2 l^2 (k^2 +k_{e}^{2} ) ] ,
\nonumber \\
B &=&  \frac{\cosh \alpha}{\alpha^{2}} 
+\frac{(\alpha^2 -1) \sinh \alpha}{\alpha^{3}} ,
\eea
where $\alpha = 4 l^{2} k k_e$. 
For the case of complete overlap $D=0$ the remaining coefficient has 
the simple form
\beq
C_{D=0} = 3B -4 \frac{\sinh \alpha}{\alpha} .
\eeq
In the limit of ground state mode 
functions much larger than the wavelength of light, $\alpha \gg 1$, 
the power reduces to 
\beq
P_{D=0} =  N_e N_{g} \kappa ( 8 \pi )^{3/2}
\frac{l^3}{ V } \frac{ 2 \pi }{4 (kl)^{2} }
d^{2} ( 1  - \cos \delta \theta_{+-} ) ,
\label{power short}
\eeq
where for simplicity we have assumed here, and hereafter, that 
$N_{+}=N_{-}=N_{g}$, $d_{+}=d_{-}=d$, and $k=k_{e}$.  This expression 
shows that the total spontaneously emitted power may vary substantially 
with the condensate phase difference $\delta \theta_{+-}$.  We note 
that the excited state decay rate is proportional to the total emitted 
power.  This means that the depletion of the excited state, as 
measured by direct detection of excited state atoms, could also be 
used to detect the phase difference.

The power Eq.(\ref{power short}) can be scaled to an emission rate by 
dividing by the photon energy $\hbar c k$.  Using the free space 
spontaneous emission rate 
$\gamma = d^{2} k^{3} /(3 \pi \varepsilon_{0} \hbar) $
and the numerical values preceding Eq.(\ref{C 
numerical}) gives the total atom stimulated emission rate
\bea
\gamma_{stim} &=& \gamma N_e N_{g} \frac{3}{16}
\frac{ (8 \pi)^{3/2} }{ (kl)^{2} }
\left( \frac{l^3}{ V } \right) 
( 1  - \cos \delta \theta_{+-} )
\nonumber \\
&=& \gamma N_e N_{g} (2.4 \times 10^{-5})
( 1  - \cos \delta \theta_{+-} ) .
\eea
For $\cos \delta \theta_{+-} = 0$ this is one hundred times the free space 
emission rate $\gamma N_{e}$ when $N_{g} \approx 4.2 \times 10^{6}$.

We next consider spatially separated condensates, that is with $D 
\neq 0$ in Eq.(\ref{wf}).  The coefficient $C$ of the interference term in 
Eq.(\ref{power}) for the total radiated power is then
\beq
C = \sqrt{\pi} \sum_{j=0}^{\infty}
\frac{(-1)^{j}}{j!} \left( \frac{D}{\sqrt{2} l} \right)^{2j}
\left\{ -I_{j+ {1\over2} } (\alpha) +
2 \frac{j+ {3\over2}}{\alpha^{2}} I_{j+ {3\over2}} (\alpha) \right\} ,
\label{Cseparated}
\eeq
where $I_{n}$ is the modified Bessel function of the first kind of 
order $n$.  The terms in the series proportional to $I_{j+{1\over2}}$ 
result from the overlap of the gaussian wavefunctions.  These terms 
are dominant for wavefunction widths much bigger than the wavelength, 
$\alpha \gg 1$.  A graph of this coefficient as a function of the 
ratio of condensate width to separation is shown in Fig.  2, for 
$kl = 4\pi$.  This shows that the effect of the interference on the 
total radiated power disappears once the condensates are separated.  

The physical situation is related to the Young's two slit 
interference experiment with slits of width $2l$ and separation $2D$. 
In the one dimensional case the interference pattern is the 
product of a sinc function due to the single slit diffraction and a  
cosine due to the interference
\beq
I(m) = \left\{ \frac{\sin( kDm )}{kDm} \cos( klm +\theta_{d}) 
\right\}^{2},
\label{young}
\eeq
where $m=\sin \vartheta$, $\vartheta$ is the angle of observation of 
the interference pattern, and $\theta_{d}$ is the phase difference 
between the light waves at each slit.  This phase difference 
determines the relative positions of the interference pattern and the 
diffraction pattern.  For the two slit experiment $D>l$.  However 
overlapping condensates have $D<l$.  Consider the effect of varying 
the source phase difference $\theta_{d}$ in this case.  Since the 
interference maxima are further apart than the sinc function zeros, 
Eq.(\ref{young}) shows that the interference maxima can shift in and 
out of under the maxima of the sinc function as $\theta_{d}$ varies.  
This is the physical origin of the modulation of the total power by 
the condensate phase difference.  When $D>l$ this does not happen 
because the interference maxima are closer together.  Hence when one 
maximum shifts out another shifts in to replace it and the total power 
is approximately constant.  The overall effect is well approximated by 
considering the effect on the total power to arise from the 
overlapping portions of the condensates.

As a final example we consider the polarization dependence of the 
radiation for another geometry.  The positive frequency part of the 
electric field with polarization $\ev_i$, perpendicular to ${\bf n}$, 
is the scalar product $\hat{\Ev}^{+} (\rv) \cdot \ev_{i}^{*}$, with 
$\hat{\Ev}^{+} (\rv)$ given by Eq.(\ref{electric field}).  The 
intensity of the polarised field is given by similar expressions to 
Eq.(\ref{intensity}) and Eq.(\ref{intensity nume}) with $\Dcv_g$ 
replaced by $\Dcv_g \cdot \ev_{i}^{*} = \dv_g \cdot \ev_{i}^{*}$.  Let 
the quantization, observation, and excited atom propagation directions 
all be $z$.  With this geometry the transitions are distinguishable by 
the polarization of the emitted photon and the total intensity does 
not depend on the relative condensate phase.  However the intensity in 
particular linear polarizations does depend on the relative phase, as 
we now show.  Define the linear polarization vectors $\ev_{\beta} = 
\ev_x \cos \beta +\ev_y \sin \beta $.  Then the intensity in 
polarization $\ev_{\beta}$ is %
\beq
I_{\beta} = N_{e} N_{g} \frac{d^{2}}{2} \frac{\kappa}{|\rv|^{2}}
\left(
| C_{+} |^2 + | C_{-} |^2  - 
\mbox{Re} \{ C^{*}_{+} C_{-} 
\exp[ i ( 2 \beta +\delta \theta_{+-} ) ]
\} \right) .
\label{pol intensity 01}
\eeq
This intensity is polarization dependent due to the final term, which 
also depends on the ground state condensate phase difference.  This 
polarization dependence is diagnostic of broken symmetry in the form 
of a relative phase between the ground state condensates.  When the 
relative phases of the transition dipoles of different atoms are 
independent there is no such polarization dependence. Independence is 
impossible for condensed atoms, which are all in the same state. 
However for non-condensed atoms independence arises through collisions or 
other dephasing perturbations.

We summarise our results by listing measurable quantities that 
may depend on the ground state condensate phase difference: the 
intensities of particular electromagnetic field modes, the total 
radiated power, and the total transition rate.  The physical origin of 
these effects is the indistinguishability of the final states.

An experiment relying on {\em atom} stimulated emission into two 
condensates will be more difficult than the corresponding single 
condensate experiment.  However our numerical estimates, and the recent 
experimental interest in two condensates in the same trap \cite{Myatt,Andrews}, 
give cause for optimism concerning the feasibility of observing the 
effect of spontaneously broken phase symmetry on spontaneous emission.

This work was supported by the Marsden Fund of the Royal Society of
New Zealand, The University of Auckland Research Fund and The New 
Zealand Lottery Grants Board.

\begin{figure}
\caption{Schematic diagram of an experiment investigating the 
condensate phase dependence of spontaneously emitted light.  The solid 
ellipses represent the ground state condensates and the dashed ellipse 
represents the excited mode.  The excited mode has momentum $\hbar k$ 
in the $x$ direction.  The dipole moments are quantized in the $z$ 
direction and are shown as arrows and directed circles.  Note that 
when viewed from the $x$ direction both dipoles appear to oscillate 
linearly.}
\label{fig:1}
\end{figure}

\begin{figure}
\caption{Graph of the interference coefficient $C$, 
Eq.(\protect\ref{Cseparated}), as a function of condensate separation 
$D$.  Parameters are $kl=4\pi$ and $\alpha=4(4\pi)^{2}$, corresponding to 
condensates larger than a wavelength in size.}
\label{fig:2}
\end{figure}

\end{document}